\documentclass[pra,aps,twocolumn,superscriptaddress,showpacs,showkeys]{revtex4}
\usepackage{graphicx}
\usepackage{amsmath}
\usepackage{color}
\usepackage{amssymb}

\begin{document}
\title{ Exact Quantum Correlations of Conjugate Variables From Joint Quadrature Measurements }
\author{S. M. Roy}
\affiliation{HBCSE,Tata Institute of Fundamental Research, Mumbai}
\date{\today}
\begin{abstract}
We demonstrate that for two canonically conjugate operators $\hat{ q},\hat {p} $ ,the global correlation $\langle 
\hat{q} \hat {p} + \hat{p} \hat {q} \rangle -2 \langle \hat{ q}\rangle \langle \hat {p}\rangle$, and the local 
correlations $\langle \hat{q}  \rangle (p) - \langle \hat{ q}\rangle$ 
and   $\langle \hat{p}  \rangle (q)-\langle \hat {p}\rangle$ can be measured exactly by Von Neumann-Arthurs-Kelly 
joint quadrature measurements .
These correlations provide a sensitive experimental test of quantum phase space probabilities quite distinct 
from the probability densities of $ q,p $. E.g. for EPR states, and entangled generalized coherent states, 
phase space probabilities which reproduce the correct position and momentum probability densities have to be 
modified to reproduce these correlations as well. 
\end{abstract}
\maketitle
% May skip begin and end figure options for lectures
% Caption works in the (begin and end) figure environment only
%\begin{figure}[ht]
%\begin{center}
% Use LaTeX and DVItoPDF while using EPS files
% \includegraphics[width=.75\columnwidth]{position_density.eps}
% Use PDFLaTeX while using JPG, PNG, PDF image files
%\includegraphics[width=.75\columnwidth]{W3.jpg}
%\caption{Position density}
%\label{fig:pos_dens}
%\end{center}
%\end{figure}
%\begin{center}
 %\includegraphics[width=.75\columnwidth]{neumann.jpg}
%\end{center}

% May skip begin and end figure options for lectures
% Caption works in the (begin and end) figure environment only
%\begin{figure}[ht]
%\begin{center}
% Use LaTeX and DVItoPDF while using EPS files
% \includegraphics[width=.75\columnwidth]{position_density.eps}
% Use PDFLaTeX while using JPG, PNG, PDF image files
%\includegraphics[width=.75\columnwidth]{W3.jpg}
%\caption{Position density}
%\label{fig:pos_dens}
%\end{center}
%\end{figure}
{\bf Von Neumann-Arthurs-Kelly Joint Measurements of Conjugate Variables in Quantum Mechanics }. 
Correlations between conjugate observables, being rather different from Bell type correlations \cite{Bell} 
among commuting observables, is a largely unexplored area with possible fundamental importance.  
We present a method for exact measurement of local and global correlations between conjugate 
observables in quantum mechanics. We use the Von Neumann-Arthurs-Kelly Joint Measurements, realizable as 
heterodyne measurements in quantum optics. As a first application
these correlations are used to experimentally test proposed phase space probabilities and to constrain 
costruction of such probabilities so that they reproduce not only the quantum probability densities 
of conjugate observables but also their correlation.

Von Neumann \cite{Von Neumann} not only proposed a model Hamiltonian for accurate measurement of 
a single observable $\hat{q}$ but 
also noted that joint approximate measurements of canonically conjugate observables $\hat{ q},\hat {p} $
 were possible with accuracy limited by the uncertainty principle . Arthurs and Kelly \cite{A-K} 
generalized the Von Neumann Hamiltonian to realize such joint measurements and deduce a measurement 
uncertainty relation which was soon proved with full generality by Arthurs and Goodman  \cite{A-G}.
Braunstein, Caves and Milburn \cite{Braunstein} analyzed the optimum initial conditions and deduced  
that the optimum Arthurs-Kelly joint distribution for  $\hat{ q},\hat {p} $ is just the Husimi $Q$ function 
\cite{Husimi}. Stenholm \cite{Stenholm} proposed a realizable quantum optics Hamiltonian for such measurements.
The various facets of the uncertainty relations so revealed have been reviewed by  Busch, Heinonen and 
Lahti \cite{Busch}. The most satisfying thing is that Heterodyne measurements in quantum optics \cite{Yuen} 
can practically realize joint measurements of any pair of rotated conjugate quadratures, $(a \exp{(-i\theta ) }/
\sqrt{2} +h.c.,-ia \exp{(-i\theta ) }/\sqrt{2} +h.c.)$
where $\hat{ q},\hat {p} $ are given in terms of the photon annihilation and creation operators $a,a^{\dagger}$,
$$ \hat{ q}= (a+ a^{\dagger} )/\sqrt{2},\hat{ p}= (a- a^{\dagger} )/(i\sqrt{2}).$$

{\bf Arthurs-Kelly Results}. Their idea is that the 
system (with position and momentum operators $\hat{ q},\hat {p} $) interacts with an apparatus which has two commuting 
observables $x_1,x_2$ and approximate values of system position and momentum are extracted from accurate 
joint observation of $x_1, x_2$. The  Von Neumann-Arthurs-Kelly interaction during the time interval $(t_0,t_0+T)$ is, 
\begin{equation}
H=K (\hat{ q} \hat {p}_1 +\hat {p} \hat {p}_2),
\end{equation}
where $\hat {p}_1, \hat {p}_2$ are canonical conjugates of $x_1,x_2$ respectively, the coupling $K$ is large, 
and $T$ is small, with $KT=1$. 
During interaction time, $H$ is so strong that the free Hamiltonians of the system and 
apparatus are neglected. Arthurs and Kelly start with the system-apparatus initial state,
\begin{equation}
 \psi (q,x_1,x_2,t_0)= \phi (q) \chi_1 (x_1) \chi_2 (x_2) 
\end{equation}
where, $\phi (q) $ is the system state and the apparatus state is given by,
\begin{equation}
\chi_1 (x_1) = \pi^{-1/4} b^{-1/2} \exp (-x_1^2 /(2 b^2)),
\end{equation}
\begin{equation}
\chi_2 (x_2) = \pi^{-1/4} (2 b)^{1/2} \exp (-2 b^2 x_2^2 ) ,
\end{equation}
and $b/\sqrt{2}$ is the uncertainty of $x_1$ in the initial apparatus state. They solve the Schr$\ddot{o}$dinger equation 
exactly and obtain the final joint probability density of the apparatus variables to be 
just the Husimi function \cite{Husimi},
\begin{equation}
P(x_1,x_2) = \left \lvert \langle \phi_{b,x_1,x_2} | \phi \rangle \right \rvert ^2 / (2 \pi),
\end{equation}
where 
\begin{equation}
 \phi_{b,x_1,x_2}(q) = (2\pi b^2)^{-1/4} \exp (iqx_2-(x_1-q)^2 /(4 b^2))
\end{equation}
is a minimum uncertainty system state centred at $q=x_1, p=x_2$. Note that for any value of $b$, 
$\langle x_ 1\rangle= \langle \hat{q}\rangle $, and $\langle x_ 2\rangle= \langle \hat{p}\rangle $, but 
the dispersions in $x_1, x_2$ are larger than those for the corresponding system variables $q,p$,
\begin{equation}
 (\Delta x_1 )^2 = (\Delta q )^2 + b^2, \: (\Delta x_2 )^2 = (\Delta p )^2 + \frac {1}{4 b^2};
\end{equation} 
they obey the ``measurement or noise '' uncertainty relation , (units $\hbar =1$ ),
\begin{equation}
 \Delta x_1 \Delta x_2  \geq 1
\end{equation} 
obtained by varying $b$. Here the minimum uncertainty is twice the usual `` preparation uncertainty''.
Arthurs and Goodman \cite{A-G} gave a beautiful proof of this fundamental uncertainty relation, 
independent of any particular choice of the Hamiltonian. Further,for the $x_1$ distribution to 
approximate $q$ distribution closely, we need $b \ll \Delta q $; for the
$x_2$ distribution to approximate $p$ distribution closely, we need $b \gg (\Delta p)^{-1} $ . 
\begin{eqnarray}
 P_1(x_1)\equiv \int P(x_1,x_2) d x_2 = (2\pi)^{-1/2} b^{-1} \nonumber\\
 \int dq \vert  \phi (q) \vert ^2 \exp {(-(x_1-q)^2 /(2 b^2))} \:
\rightarrow _{b \rightarrow 0} \vert  \phi (x_1) \vert ^2 ,
\end{eqnarray}
\begin{eqnarray}
 P_2(x_2)\equiv \int P(x_1,x_2) d x_1 = (2\pi)^{1/2} b \nonumber\\
 \int dp \vert  \tilde {\phi} (p) \vert ^2 \exp {(-(x_2-p)^2 (2 b^2))} \:
\rightarrow _{b \rightarrow \infty} \vert  \phi (x_2) \vert ^2 ,
\end{eqnarray}
where $ \tilde {\phi} (p)$ denotes the Fourier transform of $\phi (q)  $.

{\bf Exact Measurement of Quantum Correlations Between Conjugate Variables.}
The above 
equations show that the exact position and momentum probability densities of the system are 
recovered by the Arthurs-Kelly measurement in the limits $ b \rightarrow 0$ and 
$b \rightarrow \infty $ respectively,i.e. in two experiments with very different initial apparatus states.
It is a pleasant surprise that the joint measurement can nevertheless give local and global 
correlations between $\hat{ q}$,and $\hat {p} $ exactly. We define,
\begin{equation}
 \langle \hat{p} \rangle (q)\equiv \frac {\langle \Lambda (q) \hat{p} + \hat{p} \Lambda (q) \rangle } 
{2 \langle \Lambda (q)\rangle };\:\langle \hat{q} \rangle (p)\equiv \frac {\langle \Lambda (p) \hat{q} + \hat{q} \Lambda (p) \rangle } 
{2 \langle \Lambda (p)\rangle },
\end{equation}
 where $\langle A \rangle $ denotes the quantum expectation value of a self-adjoint operator $A$, and the 
projection operators $\Lambda (q),\Lambda (p) $ are defined by,
\begin{equation}
 \Lambda (q) = |q \rangle \langle q|, \:\Lambda (p) = |p \rangle \langle p| .
\end{equation}
For a pure state $ |\phi \rangle$ we have the explicit expressions,
\begin{equation}
 \langle \hat{p} \rangle (q) =\frac{Re (\phi ^* (q) (-i)\partial \phi (q)/\partial q ) } {\vert \phi (q)\vert ^2 },
\end{equation}
\begin{equation}
 \langle \hat{q} \rangle (p) =\frac{Re (\tilde {\phi} ^* (p) (i)\partial \tilde {\phi} (p)/\partial p ) }
 {\vert \tilde {\phi} (p)\vert ^2 }.
\end{equation}
We shall see that the local correlations $\langle \hat{p} \rangle (q) -\langle \hat{p} \rangle $,and 
 $\langle \hat{q} \rangle (p) - \langle \hat{q} \rangle$ can be measured exactly for 
arbitrary $q$ and $p$ respectively for appropriate values of $b$. The global correlation 
$\langle \hat{q} \hat {p} + \hat{p} \hat {q} \rangle -2 \langle \hat{q} \rangle \langle \hat{p} \rangle$ is in fact 
exactly measurable for any value of $b$.

For the Arthurs-Kelly measurement we define as for a classical distribution,
\begin{equation}
 \langle x_2 \rangle _{A-K }(x_1) \equiv \int  x_2 P(x_1,x_2) d x_2 / P_1(x_1),
\end{equation}
\begin{equation}
 \langle x_1 \rangle _{A-K }(x_2) \equiv \int  x_1 P(x_1,x_2) d x_1 / P_2(x_2),
\end{equation}
\begin{equation}
 \langle x_1 x_2 \rangle _{A-K } \equiv \int  x_1 x_2 P(x_1,x_2) d x_1 d x_2 \:.
\end{equation}
Substituting the value of  $P(x_1,x_2)$, and doing the integral over $x_2$ we obtain,
 \begin{eqnarray}
   \int  x_2 P(x_1,x_2) d x_2 =(b \sqrt{2 \pi})^{-1} \int dq dq'\phi (q) \phi ^* (q') \nonumber \\
\exp{(-\frac{(x_1-q)^2 +(x_1-q')^2 } {4 b^2} ) }i \frac {\partial \delta (q-q')}{\partial q} \nonumber\\
= Re  \int \frac{dq}{b \sqrt{2 \pi} } \exp{(-\frac{(x_1-q)^2 } {2 b^2} ) }\phi ^* (q) (-i)
\frac {\partial \phi (q)}{\partial q} 
 \end{eqnarray}
where $ \delta (q-q')$ is the Dirac delta function. Similarly, we obtain,
\begin{eqnarray}
   \int  x_1 P(x_1,x_2) d x_1 = b \sqrt{2 /\pi}  \nonumber \\
 Re  \int dp\exp{(-2 b^2 (x_2-p)^2  )} \tilde{\phi} ^* (p) i
\frac {\partial \tilde{\phi} (p)}{\partial p}. 
 \end{eqnarray}
Taking the limits of $b$ going to $0$ and $\infty$  yield respectively,
\begin{equation}
 \langle x_2 \rangle _{A-K }(x_1) \rightarrow _{b \rightarrow 0} \langle \hat{p} \rangle (q=x_1),
\end{equation}
\begin{equation}
 \langle x_1 \rangle _{A-K }(x_2) \rightarrow _{b \rightarrow \infty} \langle \hat{q} \rangle (p=x_2).
\end{equation}
Thus we have proved that the quantum position probability density and the local correlation 
$\langle \hat{p} \rangle (q) -\langle \hat{p} \rangle $ can be measured exactly with the initial condition 
$ b \rightarrow 0$; the quantum momentum probability density and the local correlation 
$\langle \hat{q} \rangle (p)-\langle \hat{q} \rangle$ can be 
measured exactly with the very different initial condition $b \rightarrow \infty $.
A similar calculation shows that for any value of $b$,
\begin{equation}
 \langle 2 x_1 x_2 \rangle _{A-K } = \langle \hat{q}\hat{p} + \hat{p}\hat{q} \rangle ,
\end{equation}
the global correlation is exactly measured in the Arthurs-Kelly (A-K) measurement.
Thus, the A-K measurements with $ b \rightarrow 0$ and $b \rightarrow \infty $ equip us with exact probability densities 
of position and momentum as well as their exact local and global correlations.

{\bf Experimental test of phase space probabilities by correlation measurements}. 
We demonstrate that exact measurement of the correlations is a valuable tool to discriminate between various phase space 
probability densities which may give exactly the same position and momentum probability densities. 
The tremendous progress initiated by research on Bell inequalities and  
quantum contextuality \cite{Bell}, and their extension to phase space \cite{Auberson}
teaches us that in $2N$ dimensional phase space, a positive density can have a 
maximum of $N+1$ marginals reproducing quantum probability densities  for arbitrary states.
(E.g. for $N=2$, probability densities of $(q_1, q_2), (p_1, q_2), (p_1, p_2)$ can be reproduced.)
Of course we know that all marginals of Wigner's quasi-probability distribution \cite{Wigner} 
agree with the corresponding quantum probablities for the state $|\phi \rangle $.But we shall only cosider  
positive densities as candidates for a probability interpretation. De Broglie and Bohm \cite{dBB} proposed 
a positive phase space density which reproduces the quantum position probability density but 
fails to agree with the quantum momentum probability density \cite{Takabayasi}.
The most general positive densities with two marginals reproducing quantum position and momentum 
probabilities \cite{Cohen-Z}, and with $N+1$ marginals reproducing the corresponding quantum probabilites 
are also known \cite{Auberson}. Roy and Singh \cite{Roy-Singh}  built a new causal
quantum mechanics symmetric in $q,p$ in which the phase space density obeys positivity and 
the marginal conditions on momentum and position probabilities . For example for $N=1$, the two densities
 \begin{eqnarray}
\rho _{\epsilon }(q, p) = |\phi (q)|^2 |\tilde\phi (p)|^2 \nonumber \\
 \delta \left(\int^p_{-\infty} dp' |\tilde\phi (p')|^2 - 
\int^{\epsilon q}_{-\infty} dq' |\phi (\epsilon q', t)|^2 \right) ,
\end{eqnarray}
where $\epsilon = \pm 1 $ clearly reproduce the quantum position and momentum probabilities as marginals.
\begin{equation}
\int \rho _{\epsilon}(q, p) dp = |\phi (q)|^2; \int \rho _{\epsilon}(q, p) dq=| \tilde \phi (p) |^2 .
\end{equation}
To demonstrate the discriminatory power of the quantum correlation measurements we shall use them in several 
concrete examples to test these two phase space densities ( for $\epsilon = \pm 1 $ ), as well as the 
correlationless phase space density $|\phi (q)|^2 |\tilde\phi (p)|^2 $, all of which reproduce 
quantum $q,p$ probability densities. 
   
(i) Free particle spreading wave packets for non-relativistic particle of mass $m$. 
At the time $t_0$ of the A-K measurement, let 
\begin{eqnarray}
 \tilde{\phi} (p) = (\pi \alpha )^{-1/4} exp [-\frac {(p-\beta )^2 } {2\alpha } -it_0\frac{p^2 } {2m } ],\nonumber \\
 (\Delta p)^2 = \frac{\alpha}{2}, (\Delta q)^2=\frac{1+ (\alpha t_0/m)^2 } {2\alpha }.
\end{eqnarray}
The Roy-Singh $q,p$ symmetric causal quantum mechanics gives, for $\epsilon = \pm 1 $,
\begin{equation}
\langle p \rangle (q)_{\pm }- \langle \hat{p} \rangle = \pm \frac{\Delta p } { \Delta q}(q-\beta t_0/m),
\end{equation}
whereas  the Arthurs-Kelly correlation is,
\begin{equation}
 \langle x_2 \rangle (x_1)_{A-K} -\langle \hat{p} \rangle= 
\frac {\sqrt{(\Delta q \Delta p)^2-1/4 } } {(\Delta q)^2 +b^2 } (x_1-\beta t_0/m),
\end{equation}
and it's limit $b \rightarrow 0 $ is the true quantum correlation.The correlationless phase space density 
gives $0$ for the above correlation and the $\epsilon =- 1 $ case gives a negative correlation, both disagreeing 
with the quantum correlation, whereas the ratio of the correlation in the A-K measurement to 
that in the $\epsilon = 1 $ Roy-Singh causal density (see figure) 
approaches unity for $b/\Delta q \ll 1 $ and $\Delta q \Delta p \gg 1/2$; i.e. there is agreement with 
quantum mechanics only when the uncertainty product is large. 
%\begin{center}
% \includegraphics[width=.75\columnwidth]{ARUNABHA_CAUSAL_QM_TEST.PNG}
%\end{center}
\begin{figure}[ht]
\begin{center}
 \includegraphics[width=.95\columnwidth]{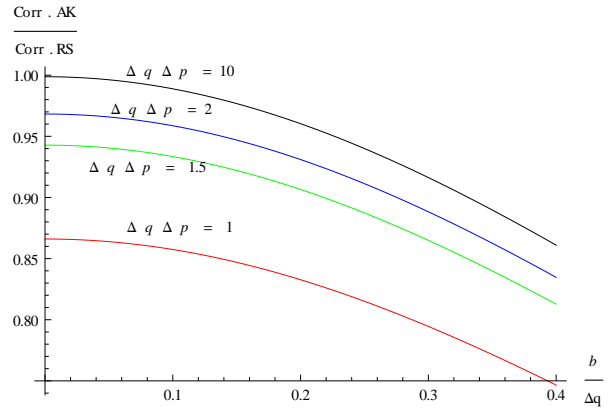}
\caption{For the free particle expanding Gaussian wave packet, the ratio of the correlation $\langle x_2\rangle(x_1)
-\langle x_2\rangle $ in the Arthurs-Kelly measurement to $\langle p\rangle(q)
-\langle p\rangle $ in the $\epsilon =1$ Roy-Singh causal phase space density is plotted for 
various values of $b/\Delta q$ and of $ \Delta q \Delta p$.The causal correlation agrees with the quantum 
correlation (i.e. the $b \rightarrow 0$ limit of the A-K correlation )only for large values of the 
uncertainty product. A convex combination of the $\epsilon =1$ and $\epsilon =-1$ causal phase space 
densities can reproduce quantum correlations exactly.(Figure computed by Arunabha S. Roy.)}
\label{fig:correlation_test}
\end{center}
\end{figure}
Similarly, for correlations at given $p$, the Roy-Singh causal quantum mechanics gives, for $\epsilon = \pm 1 $,
\begin{equation}
\langle q \rangle (p)_{\pm }- \langle \hat{q} \rangle = \pm \frac{\Delta q } { \Delta p}(p-\beta ),
\end{equation}
which agrees only for $\epsilon = 1 $ and only for large $ \Delta q \Delta p$ with the quantum correlation which is 
the $b \rightarrow \infty$ limit of the Arthurs-Kelly correlation ,
\begin{equation}
 \langle x_1 \rangle (x_2)_{A-K} -\langle \hat{q} \rangle= 
\frac {\sqrt{(\Delta q \Delta p)^2-1/4 } } {(\Delta p)^2 +(4 b^2)^{-1} } (x_2-\beta ).
\end{equation}
For the global correlation, the Roy-Singh causal quantum mechanics with $\epsilon =\pm$ gives,
\begin{equation}
 \langle 2qp \rangle _{\pm}-2\langle q \rangle \langle p \rangle= \pm 2 \Delta q \Delta p,
\end{equation}
of which only the $\epsilon =1 $ correlation agrees with the quantum correlation,
\begin{equation}
 \langle \hat{q} \hat {p} + \hat{p} \hat {q} \rangle -2 \langle \hat{q} \rangle \langle \hat{p} \rangle 
= \sqrt {(2\Delta q \Delta p)^2-1 },
\end{equation}
provided that $ 2 \Delta q \Delta p \gg 1$.

(ii) Generalized coherent states of light \cite{Roy-Singh82}.These are 
displaced $n $-th excited states of the oscillator of frequency $\omega $,
\begin{equation}
 \phi_{n,\alpha} (q,t_0) =\langle q- \bar{q }\vert n \rangle \exp{(-i\omega t_0 (n+\frac{1}{2}) +i\bar{p }
(q-\frac{\bar{p }}{2} ) }, 
\end{equation}
where $\alpha=A\exp{(-i (\omega t_0 + \theta) ) }$, $\bar{q }\equiv Re \alpha, 
\bar{p }\equiv Im \alpha$, and $A,\theta$ are real constants. Here the
 Roy-Singh causal quantum mechanics with $\epsilon =\pm 1$ gives,
\begin{equation}
 \langle p \rangle (q)_{\pm }- \langle \hat{p} \rangle = \pm (q-\bar{q}),\:
\langle q \rangle (p)_{\pm }- \langle \hat{q} \rangle = \pm (p-\bar{p}),
\end{equation}
\begin{equation}
 \langle 2qp \rangle _{\pm}-2\langle q \rangle \langle p \rangle= \pm (2n+1 ).
\end{equation}
In contrast quantum mechanics gives zero for the above three correlations and thus agrees with the correlationless 
phase space density. 
  
{\bf Phase space probabilities reproducing quantum position and momentum probabilities and correlations exactly}.
Surprisingly, in both the examples considered above, convex combinations of the Roy-Singh 
phase space densities with $\epsilon =\pm 1 $,
 \begin{equation}
 \rho (q, p)_{C} =\lambda _+ \rho _+(q, p)  + \lambda _- \rho _-(q, p),
\end{equation}
\begin{equation}
 0\leq \lambda _\pm \leq 1, \: \lambda _+ +\lambda _- = 1.
\end{equation}
where the state dependent constants $ \lambda _\pm $ are chosen to reproduce the quantum global correlation
$\langle \hat{q} \hat {p} + \hat{p} \hat {q} \rangle -2 \langle \hat{q} \rangle \langle \hat{p} \rangle $
yield local correlations also equal to the corresponding quantum local correlations. Explicitly, in cases (i) and (ii)
of Gaussian packets and generalized coherent states,
\begin{equation}
(i)\: \lambda _{\pm }= \frac{1 } {2 } \pm \frac{1 } {2 }\sqrt{1-(2\Delta q \Delta p )^{-2 } } ,(ii)\:\lambda _{\pm }=1/2.
\end{equation}

{\bf EPR states}. A normalizable version of the original EPR state \cite{EPR} $ \vert q_1-q_2=q_0\rangle 
\vert p_1+p_2=P_0\rangle$ of two particles is,
\begin{equation}
 \phi (q_1-q_2,p_1+p_2)= \phi _1 (q_1-q_2) \tilde{\phi}_2 (p_1+p_2), 
\end{equation}
where the individual Gaussian wave functions,
\begin{equation}
\phi _1 (q_1-q_2)= (\pi \alpha _1 )^{-1/4} \exp {(-\frac {( q_1-q_2 -q_0)^2 } {2\alpha _1 })},
\end{equation}
\begin{equation}
\tilde{\phi} _2 (p_1+p_2)= (\pi \alpha _2 )^{-1/4} \exp {(-\frac {(  p_1+p_2 -P_0)^2 } {2\alpha _2 })}
\end{equation}
are sharply peaked at  $q_1-q_2 -q_0 =0$ and $ p_1+p_2 -P_0 =0 $ respectively in the limits
$\alpha_1 \rightarrow 0, \alpha_2 \rightarrow 0$. We now construct the phase space density,
\begin{equation}
 \rho _{1C}(q_1-q_2,\frac{(p_1-p_2)}{2})\rho _{2C}(\frac{(q_1+q_2)}{2},(p_1+p_2)),
\end{equation}
 with the two factors $\rho _{1C} $ and $\rho _{2C} $ made to fit respectively the $(q_1-q_2,(p_1-p_2)/2)$ 
 and $ (q_1+q_2)/2,(p_1+p_2))$ correlations in the Gaussian states $\phi _1, \phi_2$  using convex combinations 
of the Roy-Singh phase space densities described above. This phase space density reproduces exactly, the above quantum 
correlations as well as quantum  joint probability densities of the four commuting pairs of variables 
$ q_1-q_2,(q_1+q_2)/2 ; q_1-q_2,p_1+p_2 ;(q_1+q_2)/2,p_1-p_2; p_1+p_2,(p_1-p_2)/2$.

{\bf Entangled generalized coherent states}.For two modes of light with the same frequency an exact solution of the 
Schr$\ddot{o}$dinger equation at time $t_0$ is the entangled generalized coherent state ,
\begin{equation}
 \phi_{m,\alpha} ((q_1+q_2)/\sqrt{2},t_0)\phi_{n,\beta} ((q_1-q_2)/\sqrt{2},t_0),
\end{equation}
where $m,n$ are integers, $\alpha,\beta $ complex constants and the factors 
$\phi_{m,\alpha},\phi_{n,\beta} $ are generalized coherent states
defined before . A phase space probability reproducing the relevant quantum correlations and probabilities exactly 
is ,  
\begin{equation}
\rho _{mC}(\frac{(q_1+q_2)}{\sqrt{2}},\frac{(p_1+p_2)}{\sqrt{2}})\rho _{nC}
(\frac{(q_1-q_2)}{\sqrt{2}},\frac{(p_1-p_2)}{\sqrt{2}})
\end{equation}
where $ \rho _{mC}, \rho _{nC}$ are arithmetic means of the $\epsilon =\pm$ Roy-Singh phase space densities 
for the states $ \phi_{m,\alpha},\phi_{n,\beta}$ respectively.

{\bf Future directions}. The central point is the exact measurability of local and global correlations 
between conjugate observables. Actual joint quadrature measurements to test their correlations will be
very interesting. The Arthurs-Kelly joint measurements 
and hence the possibilities of exact measurements of quantum correlations between conjugate 
variables can be generalized to $2N$-dimensional phase space. An interesting question is triggered by the 
success in exact reproduction of chosen quantum correlations and probabilities in the special states 
(including entangled states) considered. Can we construct phase space probabilities reproducing quantum 
position and momentum probabilities and their correlations exactly for every quantum state ?

I wish to thank Arunabha S. Roy for the crucial figure on experimental test of correlations , Virendra Singh for 
many intensive discussions and the Indian National Science Academy for financial support.

%\end{enumerate}


\begin{thebibliography}{99}
\bibitem {Bell}
J.S. Bell, {\it Physics} \underbar{1}, 195
(1964);  A.M. Gleason,  {\it J. Math. \& Mech.} \underbar{6}, 885
(1957); S. Kochen and E.P. Specker, {\it J. Math. \& Mech.}
\underbar{17}, 59 (1967).
\bibitem {Von Neumann}
J. Von Neumann, {\it Math. Foundations of Quantum Mechanics}, Princeton University Press (1955).
\bibitem {A-K}
E. Arthurs and J. L. Kelly,Jr., {\it Bell System Tech. J.} \underbar{44},725 (1965).
\bibitem {A-G}
 E. Arthurs and M. S. Goodman,
{\it Phys. Rev. Lett.} \underbar{60},247 (1988). 
\bibitem {Braunstein}
S. L. Braunstein, C. M. Caves and G. J. Milburn,
{\it Phys. Rev.} \underbar{A43},1153 (1991).
\bibitem{Husimi}
K. Husimi, {\it Proc. Phys. Math. Soc.Japan},
\underbar{22},264 (1940). 
\bibitem{Stenholm}
 S. Stenholm, {\it Ann. Phys.} \underbar{218},233 (1992);
\bibitem{Busch}
P. Busch, T. Heinonen and P. Lahti, {\it Phys. Reports} \underbar{452},155 (2007). 
\bibitem{Yuen}
H. P. Yuen and J. H. Shapiro,{\it IEEE Trans. Inf. Theory }, \underbar{24},657 (1978);\underbar{25},179 (1979);
\underbar{26},78 (1980).
\bibitem {dBB}
L. de Broglie, ``Nonlinear Wave Mechanics, A Causal
Interpretation'', (Elsevier 1960); D. Bohm, {\it Phys. Rev.}
\underbar{85}, 166; 180 (1952); D. Bohm and J.P. Vigier, {\it
Phys. Rev.} \underbar{96}, 208 (1954);P.R. Holland,{\it The Quantum Theory of Motion}, Cambridge
University Press (1993) and {\it Foundations of Physics} \underbar{28}, 881
(1998). 
\bibitem {Takabayasi}
T. Takabayasi, {\it Prog. Theor. Phys.} \underbar{8}, 143
(1952).
 \bibitem{Auberson}
G. Auberson, G. Mahoux, S. M. Roy and V. Singh,  {\it Phys. Lett.} \underbar{A300},
327 (2002); {\it Journ. Math. Phys.} \underbar{44}, 2729-2747 (2003), and \underbar{45},4832-4854 (2004);
A. Martin and S.M. Roy, {\it Phys. Lett.} \underbar{B350},
66 (1995) ; S. M. Roy, {\it Int. J. Mod. Phys.} \underbar{14},2075 (2000);
\bibitem {Wigner}
 E. Wigner, {\it Phys. Rev.} \underbar{40}, 749 (1932).
\bibitem{Cohen-Z} 
L. Cohen and Y.I. Zaparovanny,
J. Math. Phys. \underbar{21}, 794 (1980); L. Cohen, ibid
\underbar{25}, 2402 (1984).
\bibitem {Roy-Singh}
S.M. Roy and V. Singh {\it
Mod. Phys. Lett.} \underbar{A10}, 709 (1995);
S.M. Roy and V. Singh, {\it Phys. Lett.} \underbar{A255},
201 (1999). 
\bibitem{Roy-Singh82}
S. M. Roy and V. Singh, {\it Phys. Rev.} \underbar{D25},3413 (1982); 
I. R. Senitzky,{\it Phys. Rev.} \underbar{95},1115 (1954).
\bibitem{EPR}
A. Einstein, B. Podolsky and N.Rosen, {\it Phys. Rev.}
 \underbar{47}, 777 (1935).

\end{thebibliography}
\end{document}